\documentclass[prl,aps,twocolumn,amsfonts,superscriptaddress,showpacs,floatfix]{revtex4}

\usepackage{color}
\usepackage{graphicx} 
\usepackage[tight]{subfigure}
\usepackage{multirow}
\usepackage{amsmath}
\usepackage{ulem}

\newcommand{\be}{\begin{equation}}
\newcommand{\ee}{\end{equation}}
\newcommand{\bea}{\begin{eqnarray}}
\newcommand{\eea}{\end{eqnarray}}
\newcommand{\ba}{\begin{aligned}}
\newcommand{\ea}{\end{aligned}}

\def\nn{\nonumber\\}
\def\veps{\varepsilon}

\def\fr#1{(\ref{#1})}
\def\Re{\rm Re}
\def\ov{{\cal E}}
\def\k{\kappa}
\begin{document}

%%%%%%%%%%%%%%%%%%%%%%%%%%%%%%%%%%%%%%%%%%%%%%%%%%%%%%%%%%%%%%%%%%%%%%%%%%%%%
\title{Time evolution of local observables after quenching to an integrable
  model}
%%%%%%%%%%%%%%%%%%%%%%%%%%%%%%%%%%%%%%%%%%%%%%%%%%%%%%%%%%%%%%%%%%%%%%%%%%%%%

\author{Jean-S\'ebastien Caux}
\affiliation{Institute for Theoretical Physics, University of Amsterdam, Science Park 904, Postbus 94485, 1090 GL
Amsterdam, The Netherlands}

\author{Fabian H. L. Essler}
\affiliation{The Rudolf Peierls Centre for Theoretical Physics, Oxford University, Oxford, OX1 3NP, United Kingdom}

\date{\today}

\begin{abstract}
We consider quantum quenches in integrable models. We argue that the
behaviour of local observables at late times after the quench is given
by their expectation values with respect to a single representative
Hamiltonian eigenstate. This can be viewed as a generalization of the
eigenstate thermalization hypothesis to quantum integrable models. We
present a method for constructing this representative state by means
of a generalized Thermodynamic Bethe Ansatz (GTBA). Going further, we
introduce a framework for calculating the time dependence of local
observables as they evolve towards their stationary values. As an
explicit example we consider quantum quenches in the transverse-field
Ising chain and show that previously derived results are recovered
efficiently within our framework.
\end{abstract}

%\keywords{quantum quenches, integrable models}

\maketitle
{\it Introduction.} 
Recent years have witnessed dramatic progress in the study of isolated
quantum systems out of equilibrium, in particular in systems of
optically trapped ultra-cold atomic gases. Key to these advances is
the weak coupling to the environment, which allows the realization of
essentially unitary time evolution on long time scales
\cite{uc,kww-06,tc-07,tetal-11,cetal-12,getal-11}. The experimental 
results have stimulated intense theoretical efforts aimed at answering
fundamental questions such as: Do observables relax to
time-independent values? What are the principles determining
these values? How can one describe the relaxation towards stationary
behaviour? 

There is compelling evidence that nonequilibrium time
evolution is strongly affected by dimensionality and the presence of
conservation laws. The experiments of \cite{kww-06} on
trapped ${}^{87}{\rm Rb}$ atoms established that three-dimensional
condensates rapidly relax to a stationary state characterized
by an effective temperature, whereas constraining the motion of atoms
to one dimension greatly reduces the relaxation rate of the momentum
distribution function. These results spurred a flurry of 
theoretical activity aimed at shedding light on the precise effects of
integrability on the nonequilibrium dynamics of many-body quantum
systems (see 
\cite{rev,gg,rdo-08,cc-07,caz-06,bs-08,rsms-08,mwnm-09,fm-10,bkl-10,bhc-10,gce-10,CEF:2011,free,EEF:2012,rf-11,sfm-11,cic-11,2010_Mossel_NJP_12,CEF1:2012,CEF2:2012,2012_Mossel_NJP_14,2012_Caux_PRL_109,2010_Barmettler_NJP_12,2012_Foini_JSTAT_P09011,2007_Gritsev_PRL_99,2009_Moeckel_AP_324,2011_Mitra_PRL_107}
and references therein). 

So far two basic paradigms have emerged
in translationally invariant models: at late times subsystems either
thermalize, i.e. are characterized by a Gibbs distribution with an effective
temperature, or they are described by a generalized Gibbs ensemble
(GGE) \cite{gg}. When the time evolution occurs under
the action of an integrable Hamiltonian, the GGE is applicable. 
Questions regarding the approach towards the steady state long after
the quench remain difficult to tackle. Short and intermediate times
can be efficiently studied by algorithms based on matrix-product
states \cite{mwnm-09,bhc-10,2010_Barmettler_NJP_12,cetal-12}, while
numerical methods based on integrability have allowed to access
arbitrary times in finite systems
\cite{2010_Mossel_NJP_12,2012_Mossel_NJP_14,2012_Caux_PRL_109}. The 
only cases which have been largely understood are noninteracting
theories such as the transverse field Ising chain
\cite{rsms-08,CEF:2011,CEF1:2012,CEF2:2012,EEF:2012,free}. 

It is our purpose here to develop an efficient framework for the
description of the out-of-equilibrium dynamics of a system evolving
under an integrable Hamiltonian $H(h)$, where $h$ is a system
parameter such as an interaction strength or a magnetic field. Our
approach applies equally to quantum spin chains and to continuum
theories like the Lieb-Liniger model. The
situation we have in mind is that of a quantum quench: a given system
is prepared in the ground state $|\Psi\rangle$ of the
short-ranged Hamiltonian $H(h_0)$, which itself may not be
integrable. At time $t=0$ the system parameter is suddenly changed
from $h_0$ to $h$, and the system evolves unitarily under $H(h)$ for
all $t>0$, i.e. $|\Psi(t)\rangle=e^{-iH(h)t}|\Psi\rangle$. Our main
focus is the calculation of the expectation values of generic, local (in
space) operators ${\cal O}$
\begin{equation}
\langle {\cal O} (t) \rangle = \frac{\langle \Psi(t)| {\cal O} | \Psi (t) \rangle}{\langle \Psi(t) | \Psi(t) \rangle}.
\label{eq:Ot1}
\end{equation}
Examples of ${\cal O}$ would be products of spin operators located in
a finite segment of a spin chain, or density or field operators in
quantum gases. 

Our main result is to show that in the thermodynamic
limit $L\to\infty$, at fixed particle density $N/L$ and for a wide
class of local  observables, the expectation value (\ref{eq:Ot1}) can
be expressed in a simple way in terms of projections onto a {\it
single} judiciously-chosen representative `saddle point' 
eigenstate $| \Phi_{s} \rangle$ of $H(h)$: 
\begin{eqnarray}
\lim_{N \rightarrow \infty} \langle {\cal O}(t) \rangle 
= \lim_{N \rightarrow \infty} \left[
\frac{\langle \Psi | {\cal O}(t) | \Phi_{s} \rangle}{2\langle \Psi
  | \Phi_{s} \rangle}
+\Phi_s\leftrightarrow\Psi\right].
%+\frac{\langle \Phi_s| {\cal O}(t) | \Psi\rangle}{2\langle \Phi_s |
%  \Psi \rangle}. 
\label{eq:Ot1a}
\end{eqnarray}
In the stationary state, i.e. the limit $t\to\infty$, the averages of
local observables are simply given by the expectation value in the
eigenstate $|\Phi_s\rangle$
\begin{eqnarray}
\lim_{t\to\infty}\lim_{N \rightarrow \infty} \langle {\cal O}(t) \rangle = \lim_{N
  \rightarrow \infty} \frac{\langle \Phi_s | {\cal O} | \Phi_{s}
  \rangle}{\langle \Phi_s | \Phi_{s} \rangle}. 
\label{stat}
\end{eqnarray}
We stress that no time averaging is involved in \fr{stat},
which can be thought of as a generalization of the eigenstate
thermalization hypothesis \cite{ETH} to local observables in
integrable models. We wish to emphasize that the representation
\fr{eq:Ot1a} offers a dramatic reduction in computational complexity
as compared to earlier approaches.

{\it GTBA approach to nonequilibrium evolution.} Let us consider a
quantum integrable model with Hamiltonian $H$ solvable by Bethe
Ansatz 
%\cite{GaudinBOOK,KorepinBOOK}
\cite{KorepinBOOK}. Let $\{|\Phi\rangle\}$ be a
complete set of eigenstates, i.e. $H | \Phi \rangle = \omega_{\Phi} | \Phi \rangle$.
The time evolution of an arbitrary initial state $|\Psi\rangle$ is
then given by
\begin{equation}
|\Psi (t) \rangle = \sum_{\Phi} e^{-\ov_{\Phi}} e^{-i \omega_{\Phi} t} | \Phi \rangle,
\label{eq:Psit}
\end{equation}
where $\ov_\Phi$ are constant, complex-valued overlaps 
\begin{equation}
\ov_{\Phi} \equiv - \ln \langle \Phi | \Psi\rangle.
\label{eq:overlaps}
\end{equation}
Substituting (\ref{eq:Psit}) into the numerator of (\ref{eq:Ot1}) gives
a spectral representation of the form
\begin{equation}
\langle\Psi|{\cal O}(t)|\Psi\rangle=\sum_{\Phi, \Phi'} e^{-\ov^*_{\Phi} - \ov_{\Phi'}}
e^{i (\omega_{\Phi} - \omega_{\Phi'}) t}
\langle \Phi | {\cal O} | \Phi' \rangle.
\label{eq:Ot2}
\end{equation}
This double sum over a full Hilbert space basis is a serious
bottleneck. To proceed further, we look to the thermodynamic limit. In the
Yang-Yang approach to equilibrium thermodynamics 
%of integrable models 
\cite{1969_Yang_JMP_10}, a summation over
states is recast as a functional integral over root densities
$\rho$ 
\begin{equation}
|\Phi \rangle \rightarrow | \rho \rangle, \hspace{5mm} \sum_{\Phi} (...) \rightarrow \int {\cal D}[\rho] e^{S_\rho} (...).
\label{TL}
\end{equation}
Here $S_\rho$ is the entropy of all states characterized by a given
root density and $(...)$ represents quantities with well-defined
thermodynamic limits. 
Using \fr{TL} once, we can formally recast \fr{eq:Ot2} in the form
\bea
\int {\cal D}[\rho] e^{S_\rho}
\sum_{\Phi}\left[ e^{-\ov^*_{\Phi} - \ov_{\rho}}
e^{i (\omega_{\Phi} - \omega_{\rho}) t}
\frac{\langle \Phi | {\cal O} | \rho \rangle}{2}
+\Phi\leftrightarrow\rho\right].
%\nn
%+\frac{1}{2}\int {\cal D}[\rho] e^{S_\rho - \ov_{\rho}^*}
%\sum_{\Phi} e^{-\ov_{\Phi}}
%e^{i (\omega_{\rho} - \omega_{\Phi}) t}
%\langle \rho | {\cal O} | \Phi \rangle.
\label{numer}
\eea
The reason for using \fr{TL} only once is that we are interested in
local operators ${\cal O}$. These have the property that $\langle \Phi
| {\cal O} | \Phi' \rangle\neq 0$  only if both $| \Phi \rangle$ and
$| \Phi'\rangle$ scale to the same distribution $\rho$ up to
microscopic differences \endnote{In a
  large, finite system of size $L$, the changes to the root
  distribution would be of order $L^{-1}$.}. 
In the thermodynamic limit the denominator in (\ref{eq:Ot1}) becomes
\be
\langle\Psi|\Psi\rangle=\int {\cal D} [\rho] e^{-2 \Re (\ov_{\rho }) + S_{\rho}}.
\label{denom}
\ee
and can be evaluated by the method of steepest descent. The right-hand
side of \fr{denom} can be viewed as the partition function of an
integrable model with ``generalized free energy''
\begin{equation}
{\cal F}_\rho \equiv 2 \Re (\ov_{\rho}) - S_{\rho}.
\label{eq:QA}
\end{equation}
Here $S_\rho$ is the usual Yang-Yang entropy of the integrable
Hamiltonian $H(h)$. In the simplest scalar case, realized e.g. in the
Lieb-Liniger model, it takes the form $S_\rho = N \int
d\lambda \left[ (\rho +  \rho^h) \ln (\rho + \rho^h) -   \rho \ln \rho
  - \rho^h \ln \rho^h   \right]$. The hole density 
$\rho^h$ is related to the particle density $\rho$ by the
thermodynamic form of the Bethe equations 
\be
\rho(\lambda) + \rho^h
(\lambda) = \frac{1}{2\pi} + \int d\lambda' K(\lambda - \lambda') \rho(\lambda'),
\ee
where $K(\lambda)$ is a known function for a given integrable model.
The first term in \fr{eq:QA} plays the role of an effective energy per
temperature and hence acts as the ``driving term'' in a generalized Thermodynamic
Bethe Ansatz (for details, see \cite{2012_Mossel_JPA_45,2012_Caux_PRL_109}).  
Since the effective overlaps (\ref{eq:overlaps}) are strictly bounded
from below, there exists a saddle-point at $\rho_{s}$, i.e.
$\frac{\delta {\cal F}_\rho}{\delta \rho}
|_{\rho_{s}} = 0$ \footnote{For simplicity, we assume the saddle-point
to be unique, and to occur in the bulk rather than at a boundary of
the Hilbert space. Generalizations are straightforward.}. In the
thermodynamic limit, fluctuations around the 
saddle point are negligible and thermodynamic averages can be
calculated with respect to the energy eigenstate characterized by
$\rho_s$. Given that the expectation values of all local integrals of
motion in this state are by construction the same as those of the
generalized Gibbs ensemble corresponding to $H(h)$ and
$|\Psi(t=0)\rangle$, the saddle-point average of local observables
precisely reproduces the GGE average in the sense of
\cite{CEF1:2012}. 
The functional integrals in \fr{numer} can be evaluated
analogously: Given that $\langle\Phi|{\cal O}|\rho\rangle$ is non-zero
only for states $\langle\Phi|$ such that $\omega_\Phi-\omega_\rho$ and
$\ov_\Phi^*+\ov_\rho$ are intensive, the first term in \fr{numer} is
dominated by the same saddle point $\rho_s$. The second term is
treated analogously. Putting everything together we obtain the
thermodynamic limit of \fr{eq:Ot1a}. In practice we consider the
theory in a large, finite volume $L$ (at fixed density $N/L$) and a
particular, representative eigenstate $|\Phi_s\rangle$ that reduces to
$|\rho_s\rangle$ in the thermodynamic limit. The corresponding
spectral representation is then 
\bea
\langle {\cal O} (t) \rangle
&=& \lim_{N\to\infty}
\sum_\Phi \bigg\{e^{\ov_{\Phi_s}^*-\ov_{\Phi}^*
+i(\omega_{\Phi}-\omega_{\Phi_s}) t}
\frac{\langle \Phi | {\cal O} | \Phi_{s} \rangle}{2}\nn
&&\qquad+
e^{\ov_{\Phi_s}-\ov_{\Phi}
-i(\omega_{\Phi}-\omega_{\Phi_s}) t}
\frac{\langle \Phi_{s}|{\cal O} | \Phi \rangle}{2}\bigg\}.
\label{eq:spcorr}
\eea
The gain in efficiency in \fr{eq:spcorr} as compared to the ``bare''
spectral representation \fr{eq:Ot2} is manifold: first of all, a single sum remains; moreover, this sum needs in practice to be carried out only over the subset of states with non-negligible matrix elements.
We note that in our line of arguments we have not had to require $t$ to be
large. We conjecture that \fr{eq:spcorr} describes the time evolution
of local observables in the thermodynamic limit, at arbitrary times
after the quench. Importantly, in the limit
$t\to\infty$ the integrals over the summation over $\Phi$ can be
carried out by a stationary phase approximation. This shows that in
the stationary state only the expectation value in $|\Phi_s\rangle$
survives in \fr{eq:spcorr}, i.e. 
\begin{equation}
\lim_{t\to\infty}\langle {\cal O} (t) \rangle
= \lim_{N\to\infty}\langle \Phi_{s}| {\cal O} | \Phi_{s} \rangle.
\label{statstate}
\end{equation}
The physical content of \fr{eq:spcorr} is as follows. In the
thermodynamic limit $\langle {\cal O} (t) \rangle$ is fully determined
by states within a basin defined by the saddle point density
$\rho_s$. The latter defines particular simultaneous
eigenstates $|\Phi_s\rangle$ of all local conservation laws (including
the Hamiltonian $H(h)$). 
%This
%can be viewed as a generalization of the eigenstate thermalization
%hypothesis \cite{ETH} to local observables in integrable systems. 
At late, finite times $\langle {\cal O} (t) \rangle$ is determined by
quantum interference effects due to ``excitations'' over the state
$|\Phi_s\rangle$.

{\it An explicit example: the Transverse Field Ising Chain.}
The Hamiltonian of the TFIC is given by
\be
H(h)=-J\sum_{j=1}^L\Bigl[\sigma_j^x\sigma_{j+1}^x+h\sigma_j^z\Bigr]\, ,
\label{Hamiltonian}
\ee
where $\sigma^\alpha_j$ are Pauli matrices at site $j$ of a one
dimensional chain and we consider $J,h>0$. At zero temperature and in
the thermodynamic limit, the TFIC exhibits ferromagnetic long-range
order along the x-direction for $h<1$, while it is in a paramagnetic
phase for $h>1$ \cite{sachdev}. The two phases are separated by a
quantum critical point in the Ising universality class. It is well
known that $H(h)$ can be diagonalized by combined Jordan-Wigner and
Bogoliubov transformations \cite{sachdev} 
\be
H(h)=\sum_{p}
\varepsilon_h(p) \left(\alpha^\dagger_p\alpha_p-\frac{1}{2}\right),
\ee
where the single-particle energy is given by
$\varepsilon_h(k)=2J\sqrt{1+h^2-2h\cos k}$.
Our quench protocol is as follows: we prepare the
system in the ground state $|\Psi\rangle$ for an initial value $h_0$
of the transverse magnetic field. At time $t=0$ we instantaneously
change the field from $h_0$ to $h$. The state of the system at times
$t>0$ is obtained by evolving with respect to the new Hamiltonian $H(h)$,
\be
|\Psi(t)\rangle=e^{-iH(h)t}|\Psi\rangle.
\ee
The reduced density matrix of a subsystem $A$ at time $t$ after the
quench is given by 
$\rho_A(t)={\rm Tr}_{\bar{A}}\rho(t)={\rm Tr}_{\bar{A}}
|\Psi(t)\rangle\langle\Psi(t)|$, in which $\bar{A}$ is
  the complement of $A$. For quenches originating in the 
paramagnetic phase, i.e. $h_0>1$, the $\mathbb{Z}_2$ symmetry of
rotations by $\pi$ around the z-axis remains unbroken and
it is possible to express $\rho_A(t)$ in the form \cite{FE:2012}
\be
\rho_A(t)=\frac{1}{2^\ell} \sum_{\mu_l=0,1}
\Bigl< \prod_{l=1}^{2\ell} a_l^{\mu_l}\Bigr>\left(\prod_{l=1}^{2\ell}
a_l^{\mu_l}\right)^\dag \propto e^{a_l W_{lm} a_m/4}\ .
\label{quad}
\ee
Here the expectation value is with respect to the state
$|\Psi(t)\rangle$ and $a_{2n}$ and $a_{2n-1}$ are Majorana fermion
operators fulfilling anticommutation relations
$\{a_j,a_k\}=2\delta_{j,k}$, which are related to the lattice spins by
a Jordan-Wigner transformation 
\be
a_{2n-1}=\Big[\prod_{m<n}\sigma^z_m\Big] \sigma^x_n\ ,\quad
a_{2n}=\Big[\prod_{m<n}\sigma^z_m\Big] \sigma^y_n \ .
\label{Majoranas}
\ee
The matrix $W$ is given by $\tanh\frac{W}2=\Gamma$ \cite{vidal}, where
\be
\label{Gamma}
\Gamma_{jk} ={\rm Tr}\left[\rho(t)a_ka_j\right]-\delta_{j,k}= -\Gamma_{kj}.
\ee
%It was demonstrated in Ref.~[\onlinecite{CEF2:2012}] that at late times
%$t\to\infty$, and in the thermodynamic limit, the nonzero matrix
%elements are 
%\be
%\Gamma_{2l-1,2l-2n}(\infty)=-i\int_{-\pi}^\pi\frac{\mathrm d k}{2\pi}e^{-i j
%  k}e^{i\theta_k}\cos\Delta_k\, ,
%\label{Gamma1}
%\ee
In the thermodynamic limit, the correlation matrix is given by \cite{CEF2:2012}
$\Gamma_{2n-1,2j-1}=\Gamma_{2j,2n}=f_{j-n}$,
$\Gamma_{2n-1,2j}=g_{n-j}$ with
\bea
g_l&=&-i\int_{-\pi}^\pi\frac{\mathrm d k}{2\pi}e^{-i kl}
\frac{h-e^{ik}}{\sqrt{1+h^2-2h\cos k}}\nn
&&\qquad\qquad\times \left[\cos\Delta_k-i\sin\Delta_k\cos(2\varepsilon_h(k)t)\right],\nn
f_l&=&\int_{-\pi}^\pi\frac{\mathrm d k}{2\pi}e^{-i
  kl}\sin\Delta_k\sin(2\varepsilon_h(k)t),
\label{Gamma1}
\eea
%\bea
%g(k)&=&-i(h-e^{ik})\frac{
%\cos\Delta_k-i\sin\Delta_k\cos(2\varepsilon_h(k)t)}{\sqrt{1+h^2-2h\cos k}},\nn
%f(k)&=&\sin\Delta_k\sin(2\varepsilon_h(k)t).
%\eea
where
$\cos\Delta_k=\frac{4J^2(1+h h_0 -(h+h_0) \cos k)}{
\varepsilon_h(k)\varepsilon_{h_0}(k)}$.
The reduced density matrix \fr{quad} is Gaussian, and hence
multi-point correlation functions are obtained by Wick's theorem.
Concomitantly all local correlation functions in the stationary state
are fully specified by the two-point averages \fr{Gamma},
\fr{Gamma1} in the limit $t\to\infty$. So far we have considered only
the case $h_0>1$. For quenches originating in the ferromagnetic phase,
i.e. $h_0<1$, the reduced density matrix $\rho_A(t)$ is not Gaussian
\cite{FE:2012}. However, as shown in \cite{CEF2:2012}, $\rho_A(\infty)$
is again given by the $t\to\infty$ limit of \fr{quad}.

{\sl Stationary Behaviour.} We will now show how to recover these
results in the GTBA framework. The simplest way to obtain the solution
of the GTBA equations for the TFIC is to note that the mode occupation
numbers constitute conserved quantities
$[\alpha^\dagger_k\alpha_k,H(h)]=0$. Hence the root density in the 
stationary state is simply given by
\be
\rho(k)=\frac{\langle\Psi|\alpha^\dagger_k\alpha_k|\Psi\rangle}{2\pi}
=\frac{1-\cos\Delta_k}{4\pi},
\label{rhok}
\ee
and the particle density is $D=\int_{-\pi}^\pi dk\ \rho(k)$.
The corresponding Hamiltonian eigenstate at density $D=2N/L$ in a large,
finite volume is then
\be
|\Phi_{s}\rangle=\prod_{j=1}^N \alpha^\dagger_{\kappa_j}\alpha^\dagger_{-\kappa_j}
|0;h\rangle\ ,
\label{fs}
\ee
where $|0;h\rangle$ is the fermionic vacuum state and the momenta
$\kappa_j>0$ are distributed according to the density \fr{rhok}, i.e. 
$\kappa_{j+1}=\kappa_j+1/[L\rho(\kappa_j)]$.
The density matrix corresponding to the state \fr{fs} is
$\rho_s=|\Phi_{s}\rangle\langle\Phi_{s}|$ and by virtue of
the product form \fr{fs} it is Gaussian. This means that it can be
represented in the form \fr{quad} and is completely determined by its
correlation matrix \fr{Gamma}. The only non-vanishing matrix elements are
\be
\big(\Gamma_s\big)_{2l-1,2l-2n}=-\frac{i}{L}\sum_{k} 
\frac{e^{-ink}(h-e^{ik})(1-2\delta_{k,\kappa_j})}{\sqrt{1+h^2-2h\cos k}}.
\label{Gamma_stat}
\ee
Turning the sum over momenta into an integral by means of the
Euler-Maclaurin sum formula, we find that $\Gamma_{s}=\Gamma(\infty)$ and hence
$\lim_{t\to\infty}\rho_A(t)=\rho_{s,A}$. This proves that the
GTBA formalism reproduces the correct stationary state for the reduced
density matrix for any finite subsystem in the thermodynamic limit,
and hence for all local correlation functions.

{\it Relaxation behaviour.} Our general formalism suggests that
the time evolution of local (in space) operators is given by
\fr{eq:Ot1a}, where the state $|\Phi_s\rangle$ is defined in the
previous paragraph.
We now demonstrate the validity of \fr{eq:Ot1a} for any
local operator ${\cal O}$ in the case where the quench originates in
the paramagnetic phase, such that the $\mathbb{Z}_2$ symmetry is
unbroken. The proof is as follows: we start by defining two density
matrices $\rho(t)=|\Psi(t)\rangle\langle\Psi(t)|$ and
$\rho_s(t)=|\Phi_s(t)\rangle\langle\Phi_s(t)|$. Crucially, both of
these density matrices are Gaussian as a Wick's theorem holds for
averages calculated with respect to both of them. The right hand side
of \fr{eq:Ot1a} can be written in the form
\be
\left[\frac{{\rm Tr}\left[\rho_{s}(t)\rho(t){\cal O}\right]}
{2\ {\rm Tr}\left[\rho_{s}(t)\rho(t)\right]}
+\rho\leftrightarrow\rho_s\right]
\equiv
{\rm Tr}\left[\hat{\rho}(t){\cal O}\right].
\ee
%where $\hat{\rho}(t)=(2\ {\rm  Tr}\rho_s(t)\rho(t))^{-1}[\rho_s(t)\rho(t)+\rho(t)\rho_2(t)]$.
Because each term in $\hat\rho$ is a product of two Gaussian density matrices, it
is Gaussian itself, and hence fully characterized by its
correlation matrix 
\be
\hat{\Gamma}_{jk}={\rm Tr}\left[\hat{\rho}(t)a_ka_j\right]-\delta_{j,k}.
\label{Gammahat}
\ee
The two-point functions in \fr{Gammahat} are easily calculated, and as
shown in the supplementary material we have
$\lim_{N\to\infty}\hat{\Gamma}(t)=\Gamma(t)$. This proves that for any
local observable ${\cal O}$ (such as products of spin operators
contained in a finite subsystem) in the thermodynamic limit 
$\lim_{N\to\infty}
{\rm Tr}\left[\hat{\rho}(t){\cal O}\right]=
{\rm Tr}\left[\rho(t){\cal O}\right]$, and establishes eqn
\fr{eq:Ot1a}. 

Our line of arguments breaks down for quenches originating in the
ferromagnetic phase, because the density matrix $\rho(t)$ is no longer
Gaussian \cite{FE:2012}. In order to verify the validity of
\fr{eq:Ot1a} in this case, we have analyzed the relaxation of
the order parameter one-point function
$\langle\Psi(t)|\sigma^x_\ell|\Psi(t)\rangle$ for quenches with
$h_0,h<1$ in the regime where the density of excitations of the
post-quench Hamiltonian $H(h)$ in the initial state is small, i.e.
$\langle\Psi|\alpha^\dagger_k\alpha_k|\Psi\rangle \ll 1$. The
result for $Jt\gg 1$ in this case is \cite{CEF1:2012}
\be
\langle\Psi(t)|\sigma^x_\ell|\Psi(t)\rangle
\approx(1-h^2)^\frac{1}{8}
\exp\left[-2t\int_0^\pi\frac{dk}{\pi}\veps'_h(k)
K^2(k)\right].
\label{1point}
\ee
This result is recovered from \fr{eq:Ot1a} in a very efficient
way as follows. Taking into account boundary conditions in a large,
finite volume, the appropriate form of \fr{eq:Ot1a} for the
order parameter expectation value is
\bea
\langle\Psi(t)|\sigma^x_\ell|\Psi(t)\rangle=
{\rm Re}\left[
\frac{{}_{\rm R}\langle \Psi(t)|\sigma^x_\ell|\Phi_s(t)\rangle_{\rm NS}}
{{}_{\rm NS}\langle \Psi|\Phi_s\rangle_{\rm NS}}\right].
\label{rep_expansion3_main}
\eea
Here R/NS correspond to periodic/antiperiodic boundary conditions on
the fermions respectively, see e.g. \cite{CEF1:2012}. As shown in
\cite{CEF1:2012}, the state ${}_{\rm R}\langle\Psi(t)|$ can be written
as a linear superposition of energy eigenstates with $n$ pairs of 
fermions ${}_{\rm R}\langle\Psi(t)|=\sum_{n=0}^{L/2}{}_{\rm
  R}\langle\Psi_n(t)|$. The late-time behaviour of
\fr{rep_expansion3_main} is determined by states with $N$ pairs, i.e.
the term with $n=N$. Retaining only this contribution, and using the
known form of matrix elements of the order parameter \cite{FFs}, one
readily obtains the result \fr{1point} by means of the techniques
developed in \cite{CEF1:2012}. More details are provided in the
supplementary material.

{\it Conclusions.}
We have argued that averages of local observables in the steady state
reached long after a quantum quench to an integrable model can be
described as expectation values with respect to a single eigenstate of
the Hamiltonian (eqns (\ref{stat}) and (\ref{statstate})). This
eigenstate can be constructed by means of a generalized Thermodynamic
Bethe Ansatz and corresponds to a saddle point of an appropriately
defined ``free energy''. Going further, we have shown 
that the time evolution of local observables is governed by states in
the vicinity of this saddle-point through eqns (\ref{eq:Ot1a}) and
\fr{eq:spcorr}. The spectral representation \fr{eq:spcorr} allows to
identify the physical mechanism underlying the relaxation for a given
observable at late times. Our approach paves the way for analyzing
quantum quenches in interacting integrable models and applications to
the Lieb-Liniger and the sine-Gordon model are in progress. An interesting
question concerns the range of initial states that can be analyzed
within our framework. In quenches of the kind we have focussed on
here, the initial state is translationally invariant and characterized
by low entanglement. Whether these are necessary
requirements and whether the regularity assumptions underlying our
GTBA analysis hold for any such states are important points for
further investigation.

\acknowledgments
This work was supported by the Foundation for Fundamental Research on Matter (FOM) and the
Netherlands Organisation for Scientific Research (NWO) (JSC), 
the EPSRC under grants EP/I032487/1 and EP/J014885/1 (FHLE) and the
National Science Foundation under grant NSF PHY11-25915 (JSC and
FHLE). We thank the KITP in Santa Barbara for hospitality.

\appendix

%%%%%%%%%%%%%%%%%%%%%%%%%%%%%%%%%%%%%
\section{\large Supplementary Material}
%%%%%%%%%%%%%%%%%%%%%%%%%%%%%%%%%%%%%
%%%%%%%%%%%%%%%%%%%%
\subsection{Calculation of the correlation matrix $\hat\Gamma$ \fr{Gammahat}}
%%%%%%%%%%%%%%%%%%%%
For quenches originating in the disordered phase the $\mathbb{Z}_2$
symmetry is unbroken and no subtleties emerge with regards to boundary
conditions in a large, finite volume $L$. In particular all momenta in
the formulas below can be taken in the NS sector,
i.e. $p_n=\frac{2\pi}{L}(n+1/2)$, where $-L/2\leq n< L/2$. 
The state $|\Psi(t)\rangle$ can be expressed in terms of eigenstates of
$H(h)$ as \cite{CEF1:2012}
\be
|\Psi(t)\rangle=
\prod_{k>0}
\frac{1+ie^{-2it\varepsilon_h(k)}K(k)\alpha^\dagger_{-k}\alpha^\dagger_k}
{\sqrt{1+K^2(k)}}|0;h\rangle,
\ee
where $K(k)=\tan\big(\Delta_k/2\big)$, $\alpha_k$ and
$\alpha^\dagger_k$ obey canonical anticommutation relations, and
$\alpha_k|0;h\rangle=0$. The time evolved state
$|\Phi_s\rangle $ \fr{fs} is 
\be
|\Phi_s(t)\rangle
=\prod_{j=1}^{N}
e^{-2i\varepsilon_h(\kappa_j)t}\alpha^\dagger_{\kappa_j}\alpha^\dagger_{-\kappa_j}
|0;h\rangle\ .
\label{fs2}
\ee
Our goal is to determine all two-point matrix elements of the form
$\langle\Psi(t)|a_ja_k|\Phi_s(t)\rangle$, where $a_j$ are the Majorana
fermion operators defined in \fr{Majoranas}. The mode expansion of the
$a_j$ is 
\bea
a_{2j-1}&=&\frac{1}{\sqrt{L}}\sum_pe^{i\frac{\theta_p}{2}-ipj}\left[\alpha_p
+\alpha^\dagger_{-p}\right],\nn
a_{2j}&=&-\frac{i}{\sqrt{L}}\sum_pe^{-i\frac{\theta_p}{2}-ipj}\left[\alpha_p
-\alpha^\dagger_{-p}\right],
\eea
where
\be
e^{i\theta_k}=\frac{h-e^{ik}}{\sqrt{1+h^2-2h\cos k}}.
\label{eq:thetak}
\ee
A straightforward calculation using the mode expansion and
$\alpha_p|0;h\rangle=0$ gives
\begin{widetext}
\bea
\frac{\langle\Psi(t)|a_1a_{2n+1}
  |\Phi_s(t)\rangle}{\langle\Psi|\Phi_s\rangle}
=\delta_{n,0}-\frac{i}{L}\sum_pK(p)e^{ipn+2it\varepsilon_h(p)}
+\frac{i}{L}\sum_{j=1}^{N} \left[e^{in\kappa_j}-e^{-in\kappa_j}\right]
\left[K(\kappa_j)e^{2it\varepsilon_h(\kappa_j)}
+\frac{e^{-2it_h(\kappa_j)}}{K(\kappa_j)}\right].
\eea
\end{widetext}
In the limit $N,L\to\infty$ with $D=2N/L$ fixed we can turn sums into
integrals using the Euler-Maclaurin sum formula
\be
\frac{1}{L}\sum_{j=1}^{N}f(\kappa_j)=\int_{0}^\pi dk\ \rho(k)f(k)+{\cal
  O}\big(L^{-1}\big),
\ee
where $\rho(k)$ is the root density \fr{rhok} characterizing the
distribution of the momenta $\{\kappa_j\}$. Using that
$K(k)=\tan(\Delta_k/2)$ we find
\bea
\lim_{N\to\infty}\frac{\langle\Psi(t)|a_1a_{2n+1}
  |\Phi_s(t)\rangle}{\langle\Psi|\Phi_s\rangle}
=\delta_{n,0}+f_{-n},
%-\int_{-\pi}^\pi\frac{dk}{2\pi}e^{inp}\sin\Delta_p\sin(2\varepsilon_h(p)t).
\label{one}
\eea
where $f_l$ is given in \fr{Gamma1}. By taking the complex conjugate
of \fr{one}, we find
\bea
\lim_{N\to\infty}\frac{\langle\Phi_s(t)|a_1a_{2n+1}
  |\Psi(t)\rangle}{\langle\Phi_s|\Psi\rangle} 
=\delta_{n,0}-f^*_{-n},
\label{one_b}
\eea
Given that $f_n$ is purely imaginary, we conclude that
\be
\hat{\Gamma}_{2n+1,1}={\rm Tr}\left[\hat{\rho}(t)a_1a_{2n+1}\right]-\delta_{n,0}=
f_{-n}.
\ee
By analogous calculations one finds that
\bea
\lim_{N\to\infty}\frac{\langle\Psi(t)|a_2a_{2n+2}
  |\Phi_s(t)\rangle}{\langle\Psi|\Phi_s\rangle}
&=&\delta_{n,0}+f_n,\nn
\lim_{N\to\infty}\frac{\langle\Psi(t)|a_{2}a_{2n+1}
  |\Phi_s(t)\rangle}{\langle\Psi|\Phi_s\rangle}
&=&g_{n}.
\label{two}
\eea
Taking the complex conjugates of eqns \fr{two} and using that both
$f_n$ and $g_n$ are purely imainary, we obtain
\bea
\hat{\Gamma}_{2n+2,2}&=&{\rm Tr}\left[\hat{\rho}(t)a_2a_{2n+2}\right]-\delta_{n,0}=
f_{n}\ ,\nn
\hat{\Gamma}_{2n+1,2}&=&{\rm Tr}\left[\hat{\rho}(t)a_2a_{2n+1}\right]=
g_{n}.
\eea
Finally, using translational invariance we conclude that
\be
\hat{\Gamma}_{ij}=\Gamma_{ij},
\ee
where $\Gamma_{ij}$ is given in \fr{Gamma}, \fr{Gamma1}.
%%%%%%%%%%%%%%%%%%%%
\subsection{Order Parameter One-Point Function}
%%%%%%%%%%%%%%%%%%%%
As we are interested in the order parameter, we
need to carefully keep track of boundary conditions. We follow the
conventions summarized in the appendix of Ref.~\cite{CEF1:2012}.
The state after the quench is given by \cite{CEF1:2012}
\bea
|\Psi(t)\rangle&=&\frac{|\Psi(t)\rangle_{\rm R}+|\Psi(t)\rangle_{\rm
    NS}}{\sqrt{2}}\ ,\nn
|\Psi(t)\rangle_{\tt a}&=&
\prod_{0<k\in{\tt a}}
\frac{1+ie^{-2it\varepsilon_h(k)}K(k)\alpha^\dagger_{-k}\alpha^\dagger_k}
{\sqrt{\big(1+K^2(k)\big)}}|0;h\rangle_{\tt a},\ \ \
\eea
where ${\tt a}={\rm R,NS}$. The appropriate form of \fr{eq:Ot1a}
reads 
\bea
\langle\Psi(t)|\sigma^x_\ell|\Psi(t)\rangle=\lim_{N\to\infty}
{\rm Re}
\left[\frac{{}_{\rm R}\langle \Psi|\sigma^x_\ell(t)|\Phi_s\rangle_{\rm NS}}
{{}_{\rm NS}\langle \Psi|\Phi_s\rangle_{\rm NS}}\right].
\label{rep_expansion3}
\eea
The state ${}_{\rm R}\langle\Psi(t)|$ can be written as a linear
superposition of energy eigenstates \cite{CEF1:2012} with $n$ pairs of
fermions 
\be
{}_{\rm R}\langle\Psi(t)|=\sum_{n=0}^{L/2}{}_{\rm R}\langle\Psi_n(t)|\ .
\ee
Inspection of the resulting matrix elements suggests that at late
times the dominant contribution to \fr{rep_expansion3} arises from
terms with $N$ pairs. The corresponding contribution is
\begin{widetext}
\be
\frac{{}_{\rm R}\langle \Psi_N(t)|\sigma^x_\ell|\Phi_s(t)\rangle_{\rm NS}}
{{}_{\rm NS}\langle \Psi|\Phi_s\rangle_{\rm NS}}=\frac{(-1)^N}{N!}
\sum_{0<p_1,\dots,p_N}\!
{}_{\rm R}\langle p_1,-p_1,\dots
p_N,-p_N|\sigma^x_\ell|
\k_1,-\k_1,\dots \k_N,-\k_N\rangle_{\rm NS}\
\prod_{j=1}^{N}
%\left[-
\frac{K(p_j)}{K(\k_j)}
%\right]
e^{2it(\veps_{p_j}-\veps_{\k_j})}\ ,
\label{NDPcontrib}
\ee
\end{widetext}
where $\veps_p=\veps_h(p)$.
The matrix element of $\sigma^x_\ell$ in \fr{NDPcontrib} is given by
\cite{CEF1:2012,FFs} 
\bea
&&\frac{(-1)^{N}2^{2N}\sqrt{\xi}}{L^{2N}}\prod_{j,l}\epsilon_{\k_j,p_l}^4c^2_{\k_j,p_l}
\prod_{j=1}^{N}\frac{\sin \k_j\sin p_j}{\veps_{\k_j}^2\veps_{p_j}^2}\nn
&&\times\ 
\prod_{j<j'}\frac{1}{\epsilon_{\k_j,\k_{j'}}^4\epsilon_{p_j,p_{j'}}^4
c_{\k_j,\k_{j'}}^2c_{p_j,p_{j'}}^2}.
\eea
Here $\xi=|1-h^2|^{1/4}$, $\epsilon_{p,q}=(\veps_p+\veps_q)/2$ and $c_{p,q}=[\cos p-\cos
  q]^{-1}$. 
As a function of $p_j$ it exhibits either simple or double poles when
$p_j\rightarrow k_m$, and the leading contributions at late times
arise from the vicinities of these singularities \cite{CEF1:2012}. We
proceed by rewriting the ``pole-factor'' as the square of a Cauchy
determinant 
\bea
{\cal P}&=&\prod_{j,l=1}^N
c^2_{\k_j,p_l}\prod_{j<j'}\frac{1}{c^2_{p_j,p_{j'}}c^2_{\k_j,\k_{j'}}}=\big(\det
C\big)^2\ ,\nn
C_{ij}&=&\frac{1}{\cos p_i-\cos \k_j}\ .
\eea
By expressing the determinant as a sum over permutations using the
standard definition, we can recast ${\cal P}$ in the form 
${\cal P}=\sum_{j=0}^{[N/2]}{\cal P}_{N-2j}$, where
${\cal P}_{k}$ has $k$ double poles when viewed as a function of
the $\cos p_r$. The most singular terms are
($x_r=\cos p_r$, $y_s=\cos k_s$)
\bea
{\cal P}_N&=&\sum_{P\in S_N}\prod_{j=1}^{N}\frac{1}{(x_j-y_{P_j})^2},\nn
{\cal P}_{N-2}&=&
-\frac{1}{2}\sum_{P\in S_N}\sum_{r\neq s}\prod_{j\neq
  r,s}^{N}\frac{1}{(x_j-y_{P_j})^2}\nn
&&\hskip -1cm\times\frac{1}{(x_r-y_{P_r})(x_r-y_{P_s})(x_s-y_{P_r})(x_s-y_{P_s})}.
\label{polefactor}
\eea
The leading contribution to \fr{NDPcontrib} at late times is obtained
by retaining only the most singular part ${\cal P}_N$
\begin{widetext}
\be
\frac{{}_{\rm R}\langle \Psi_N(t)|\sigma^x_\ell|\Phi_s(t)\rangle_{\rm NS}}
{{}_{\rm NS}\langle \Psi|\Phi_s\rangle_{\rm NS}}\approx
\frac{2^{2N}\sqrt{\xi}}{L^{2N}N!}
\sum_{0<p_1,\dots,p_N}{\cal P}_N
\prod_{j=1}^{N}\left[\frac{K(p_j)}{K(\k_j)}
\frac{\sin \k_j\sin p_j}{\veps_{\k_j}^2\veps_{p_j}^2}
e^{2it(\veps_{p_j}-\veps_{\k_j})}\right]
\prod_{j,l}\epsilon_{\k_j,p_l}^4
\prod_{j<j'}\frac{1}{\epsilon_{\k_j,\k_{j'}}^4\epsilon_{p_j,p_{j'}}^4}.
\label{NDPcontrib2}
\ee
\end{widetext}
We can now carry out all sums over $p_j$ using (D.1) of
Ref.~[\onlinecite{CEF1:2012}]. Retaining only the leading
contributions at late times (and asymptotically large $L$) we find
\bea
\frac{{}_{\rm R}\langle \Psi_N(t)|\sigma^x_\ell|\Phi_s(t)\rangle_{\rm NS}}
{{}_{\rm NS}\langle \Psi|\Phi_s\rangle_{\rm NS}}\!\!&\approx&\!\!
\sqrt{\xi}
\prod_{j=1}^{N}\left(1-\frac{4t}{L}\veps'_{\k_j}\right)\nn
\!\!&\simeq&\!\!
\sqrt{\xi}
\exp\left[-\frac{4t}{L}\sum_{j=1}^{N}\veps'_{\k_j}\right].\
\eea
Turning the sum over the $\k_j$ into an integral this becomes
\be
\frac{{}_{\rm R}\langle \Psi_N(t)|\sigma^x_\ell|\Phi_s(t)\rangle_{\rm NS}}
{{}_{\rm NS}\langle \Psi|\Phi_s\rangle_{\rm NS}}\approx
\sqrt{\xi}e^{-t\int_0^\pi\frac{dk}{\pi}\veps'_{k}
\big(1-\cos\Delta_k\big)}.
\label{final}
\ee
Observing that
\be
1-\cos\Delta_k=\frac{2K^2(k)}{1+K^2(k)},
\ee
and noting that the asymptotic result \fr{final} is purely real, we
recover the result obtained in \cite{CEF1:2012} by summing the leading
late-time divergences in the form factor expansion for the 1-point
function to all orders. We stress that the current calculation is significantly
simpler. Subleading contributions at late times can be obtained by
taking into account ${\cal P}_{N-2j}$ with $j\geq 1$. These will be
reported elsewhere.

\end{document}